\documentclass[11pt]{article}

\pdfoutput=1
\usepackage[pdftex]{color}
\usepackage{amssymb}
\usepackage{amsthm}
\usepackage{amsmath}
\usepackage{latexsym}
\usepackage{amscd}
\usepackage{graphicx}
\usepackage[pdftex, colorlinks=true, citecolor=green]{hyperref}
\usepackage{lscape}
\usepackage{setspace}
\usepackage{multirow}

\newcommand{\ds}{\displaystyle}

\newcommand{\ben}{\begin{equation}}     
\newcommand{\eeqn}{\end{equation}}
\newcommand{\bey}{\begin{eqnarray}}
\newcommand{\eey}{\end{eqnarray}}

\begin{document}

\noindent {\Large
\textbf{Ebola impact and quarantine in a network model}
}
\\
\vspace{4mm}
\noindent  Anca R\v{a}dulescu$^{*,}\footnote{ Assistant Professor, Department of Mathematics, State University of New York at New Paltz; New York, USA; Phone: (845) 257-3532; Email: radulesa@newpaltz.edu}$, Joanna Herron$^{1}$
\\
\noindent $^1$ Department of Mathematics, SUNY New Paltz, NY 12561
\\

\begin{abstract}

\noindent Much effort has been directed towards using mathematical models to understand and predict contagious disease, in particular Ebola outbreaks. Classical SIR (susceptible-infected-recovered) compartmental models capture well the dynamics of the outbreak in certain communities, and accurately describe the differences between them based on a variety of parameters. However, repeated resurgence of Ebola contagions suggests that there are components of the global disease dynamics that we don't yet fully understand and can't effectively control. \\

\noindent In order to understand the dynamics of a more widespread contagion, we placed SIR models within the framework of dynamic networks, with the communities at risk of contracting the virus acting as nonlinear systems, coupled based on a connectivity graph. We study how the effects of the disease (measured as the outbreak impact and duration) change with respect to local parameters, but also with changes in both short-range and long-range connectivity patterns in the graph. We discuss the implications of optimizing both these measures in increasingly realistic models of coupled communities.

\end{abstract}

\section{Introduction}

\subsection{Background to Ebola and modeling}

\noindent The evolution, prognosis and spreading of contagious diseases has been studied for a long time, with a variety of approaches~\cite{bailey1975mathematical,anderson1992infectious,heesterbeek2000mathematical}. While huge progress has been made in efficiently applying containment methods~\cite{haas2014quarantine,pandey2014strategies} and treatment~\cite{stroher2006progress} in most cases, Ebola remains the 21st century's taunting example of a disease which mankind does not seem prepared to handle, even at small scales~\cite{chowell2004basic,feldmann2014ebola}.

While the course of the illness is quite drastic and the recovery of an Ebola patient unlikely (even with prompt clinical intervention), it has been argued that Ebola, with a relatively small $1 \leq R_0 \leq 7$ (varying among reports of different outbreaks)~\cite{chowell2004basic,legrand2007understanding,chowell2004basic}, poses a lesser contagion threat than faster spreading diseases such as small pox ($R_0 > 7$). However, the explosive Ebola contagions that periodically resurge throughout the Globe, and in particular the recent simultaneous developments in a few countries around the world~\cite{labouba2015ebola} -- seem to suggest that there are components of the disease dynamics that we don't yet fully grasp~\cite{erickson2014preparedness}. 

Much effort has been recently directed towards understanding and predicting Ebola outbreaks with the help of mathematical modeling~\cite{legrand2007understanding,rivers2014modeling,lewnard2014dynamics,siettos2015modeling}. A large volume of the existing modeling work addresses disease dynamics via low dimensional compartmental models~\cite{satsuma2004extending,hu2011bifurcations,safi2012threshold}, deterministic or stochastic~\cite{lekone2006statistical}. These models describe well outbreak dynamics in specific communities, and often explain quite accurately the differences between outbreaks that occurred under different parameter regimes (e.g., geographical location, promptitude of clinical measures, timely removal of infected individuals from the community~\cite{astacio1996mathematical,chowell2004basic,safi2012threshold}). However, these analyses -- based on localized factors and data from remote rural areas -- may all be irrelevant in the context of utmost concern, that of a global outbreak~\cite{gostin2014ebola,briand2014international}, affecting urban areas as well as small communities, acting at multiple simultaneous foci. In order to understand the dynamics of a global contagion, predict its potential effects and most efficiently alter its course -- one rather needs to place the problem in the mathematical framework of complex systems~\cite{newman2003structure,moreno2002epidemic}. The emergence of a contagion (of Ebola, or of any other viral or bacterial infection) can be then viewed as the propagation of a perturbation in a complex network of coupled nodes. 

People are organized, as many other complex systems found in nature, as a gigantic, everchanging network, functioning at multiple spatial and temporal scales. The huge oriented graph that represents worldwide physical connectivity between individuals is highly modular, has hubs, multiple levels of communities (e.g., from families to towns to states), and is constantly changing due to a variety of factors (e.g. personal, economic, commuting patterns, long distance travel). Capturing effectively the global dynamics of epidemics in such a system goes far beyond the reach of simple compartmental models. A more accurate approach should encompass simultaneously the state of each node as well as its interconnections with other nodes~\cite{eubank2004modelling}, and study not only the effect on the system of single node dynamics (e.g., the state of one infected individual, or contagion in one community), but also the effects of altering the connectivity patterns on the systemic dynamics~\cite{barthelemy2004velocity} and on the outbreak aftermath (asymptotic behavior). 

The most traditional and well-known modeling approaches to disease spread remain the SIR-type models, which describe the progress of the contagion in a single community through transfers between three main compartments: the susceptible (S), infected (I) and recovered (R) individuals. Variations of the SIR model have been vastly explored, in conjunction with parameters estimated empirically, and in increasingly complex contexts, including factors like hospitalization, treatment plans, quarantine. However, in recent years, there has been an emerging interest in \emph{network} modeling of epidemics~\cite{eubank2004modelling,barthelemy2004velocity,keeling2005networks,read2008dynamic}, trying to quantify and better understand the effect of dynamic human interactions on the spread of disease between, as much as within communities, and looking to identify the factors (both deterministic and stochastic~\cite{kenah2007network}) most determinant of the long term outcome.

\subsection{Connectomics in models of contagious disease}

Networked dynamic systems~\cite{zelazo2008observability} have been used in a variety of fields with a focus on understanding the behavior of an ensemble of coupled dynamic nodes, be them cells, web servers, people, or nodes in an electric grid. Most recently, this type of approach is being used with success from modeling the brain (through large collaborative research efforts such as the Human Connectome Project) to modeling propagation of a virus through a network of communities~\cite{kiskowski2014three}. The world population is a system comparable with the brain: in size (billions of variables), setup (nodes coupled at different spatio-temporal scales), complexity and mobility over time.  The same methods appropriate when addressing electric and chemical signals, and their propagation in a complex network of neurons, may be successful in reaching beyond the study of individual disease dynamics, and address its propagation in a complex network of people. 

It was noted, when studying the effects of contact heterogeneity on the dynamics of communicable disease~\cite{stehle2011simulation}, that small differences in the contact networks (e.g., taking into account restructuring at a time resolution of minutes), are typically not essential when attempting to describe disease spread on a longer timescale (of several weeks, or months). On the other hand, the same studies~\cite{stehle2011simulation,smieszek2009models} emphasize the importance of including detailed information about the heterogeneity of contact duration, the rate of new contacts being identified, etc. Relatively new work, including our own, is using a combination of dynamical systems and graph theoretical methods to understand how relatively small, local perturbations in connectivity patterns may produce, if targeted at the right vulnerability points, very large effects on the state of the entire network. Along these lines, we have developed methods and measures of how likely the system is to undergo a sharp transition when perturbing local connectivity~\cite{radulescu2015nonlinear}.

When studying the spread of an extremely contagious disease, we are observing a fast diffusion and mixing process, analogous in a way with the propagation of an electric charge through the brain network during a seizure. In the case of epilepsy, one aims to understand the source of the electric surge propagating through the complex brain network, and often treat it by locating and surgically excising the focal point, thus removing the source of chaos and restoring proper function of neural mechanisms. Altering dynamics in an optimal way requires knowledge of the brain architecture~\cite{bassett2006small}, of the location and density of hubs and of rich clubs~\cite{van2011rich}, of the robust features and of the vulnerable network points~\cite{bullmore2009complex} (where local edge changes produce large effects in degree node distribution, modularity).

Similarly, one can model the propagation of a virus through a complex population network, and study how adapting the structure of the network may contribute to minimizing the global effects of the contagion. Experience has shown that, along with other important factors, timely placement of temporary travel restrictions or quarantines around local communities (and if necessary around more global structures, such as countries or continents) may be critical in isolating and extinguishing the contagion before it reaches catastrophic levels~\cite{frieden2014ebola}. One can then search for the locations where a small change in connectivity is likely to produce the largest overall effects. This goes along the lines of other new approaches in the field, which introduce the network structure as a system parameter~\cite{danon2011networks,kenah2011epidemic}, and search for optimal quarantine measures to efficiently isolate the epidemic, without adding unnecessary and unfeasible burdens on the general population.

In this paper, we consider, in conjunction with our model, two measures of outbreak effects on population: the \emph{impact}, defined as the total number of individuals affected by the outbreak (eventually either recovered or dead), and the outbreak \emph{duration}, defined as the time from the start of the outbreak (presence of the first Ebola case), until the outbreak extinguishes (less than one infected case in each community). Starting with a basic SIR model in a single, relatively small (1000 individuals) community, we construct incrementally a plausible model for Ebola spread within a network of such communities. We focus in particular on studying how our two measures of outbreak dynamics depend on the density of potential connections between these communities and on the stochastic likelihood of people to travel along these connections. The network architecture is allowed to change along the process, controlled by factors related to the current state of the outbreak, accounting for travel interdictions and quarantines typically introduced in such circumstances. Our model aims to inform on the most efficient quarantining strategies and optimal timing that would minimize life loss as well as outbreak duration.

\section{Modeling methods and results}

\subsection{Basic model}

We build upon a basic compartmental model of Ebola due to Astacio et al.~\cite{astacio1996mathematical}, which was originally assembled as a classical three-variable SIR (Susceptible-Infectious-Recovery) model. In the later iterations of the model construction, the authors introduced incubation in the model,  by means of a new (fourth) variable $E$, representing the ``latent'' population (during the virus incubation, before development of symptoms).

\begin{eqnarray}
\frac{dS}{dt} &=& - \beta S(I+qE)/N \nonumber \\
\frac{dE}{dt} &=& \beta S(I+qE)/N - \delta I \nonumber \\
\frac{dI}{dt} &=& \delta I - \gamma I \nonumber \\
\frac{dR}{dt} &=& \gamma I
\label{astacio}
\end{eqnarray}

\noindent where $S(t)$ is the susceptible population at time $t$ (i.e., everyone who had not yet contracted the disease); $E(t)$ is the latent population (individuals who have contracted the virus, but are still asymptomatic); $I(t)$ represents the infected population (showing the signs and symptoms of Ebola); $R(t)$ is the number of dead or recovered individuals (i.e., in an oversimplified view,  individuals who can no longer infect others with the Ebola virus). The infection rate $\beta S(I+qE)/N$ is proportional with the product between the number of susceptible individuals and the number of individuals carrying the virus (both latent and infected, with the latent individuals having a lesser impact, represented by the smaller weight $q<1$). The proportionality constant $\beta$ is the product between the per capita contact rate and the probability of infection after contact with an infected individual. The rate was normalized by the factor $N(t)$, representing the total population at time $t$: $N(t) = S(t)+E(t)+I(t)+R(t)$. The transfer rate from the latent to the infectious stage is a fraction $\delta$ of the number of latent individuals -- where $\delta$, the per capita infectious rate, can be measured as $1/\delta$, the average time for a latent individual to become infectious. The death/recovery rate is a fraction $\gamma$ of the infectious population, where $1/\gamma$ is the average time it takes a person to die or recover once in the infectious stage.

The parameters were validated against the disease dynamics described in a relatively well localized 1976 contagion in Yambuku, Zaire. The ranges of values, as per our reference, are listed in Table~\ref{parameters}. In our study, we worked primarily with these values; changes and extensions, based on information on post-mortem and post-recovery potential for contagion, are explored in the Discussion. In this study, however, we work under the same simplifying assumption as the one made by our reference: that neither deceased nor recovered individuals can contribute any further to the perpetuation of the disease cycle, hence are represented by a common variable, whose asymptotic value $\ds R_\infty = \lim_{t \to \infty} R(t)$ can be viewed as measuring the \emph{impact} of the outbreak (total number of people affected).

\begin{table}[h!]
\begin{center}
{\renewcommand{\arraystretch}{1.25}
\begin{tabular}{|c|c|c|c|}
\hline
 {\bf Parameter} & {\bf Range} & {\bf Value} & {\bf Units}\\ 
\hline
$1/\gamma$ & 4-10 & 7 & days \\
\hline
$1/\delta$ & 2-21 & 12 & days \\
\hline
$q$ & -- & 0.25 & -- \\
\hline
$\beta$ & -- & 0.567 & -- \\
\hline
\end{tabular}}
\end{center}
\caption{\small{{\bf Parameter values for the Yumbuku outbreak,} as per the Astacio et al. reference~\cite{astacio1996mathematical}.}}
\label{load_table_XY}
\label{parameters}
\end{table}

\subsection{Simple network model}

\noindent We first studied contagion propagation in a small, unstructured network of interconnected communities. Although this preliminary model makes a few coarse and rather unrealistic simplifying assumptions, this first stage helps understand some very basic problems and questions to address. \\

\begin{figure}[h!]
\begin{center}
\includegraphics[width=\linewidth]{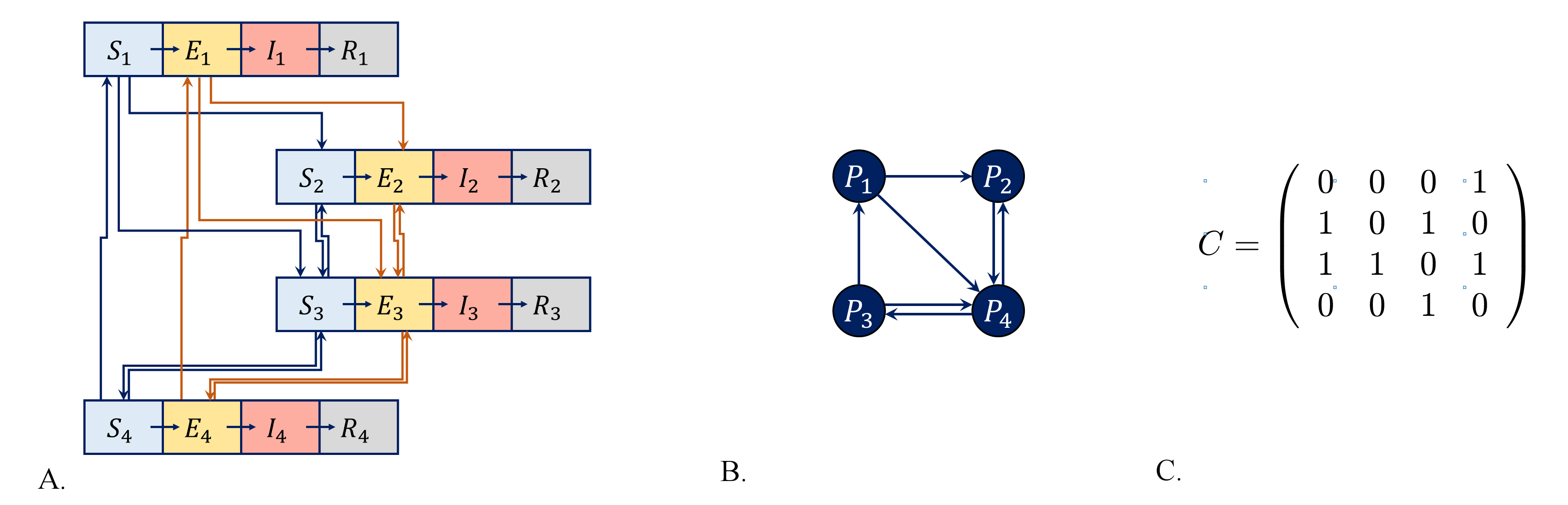}
\end{center}
\caption{\emph{\small {\bf Model network, for $n=4$ communities}, (as used in some of our later simulations).  {\bf A.} Illustration of the network, showing each population as a compartmental node, with its 4 coupled SEIR variables, and showing transfer between populations as oriented arrows from the original population to the target population. {\bf B.} Schematic representation of the same network, in which each population is viewed as a node, and the connections between populations are viewed as oriented edges. {\bf C.} Adjacency matrix $C$ corresponding to the connectivity graph described in (B).}}
\label{connect_simple}
\end{figure}

\noindent We considered a graph in which the $n$ nodes represent the interconnected communities, and the oriented edges connecting node pairs represent one-way communication between the two respective nodes. Each node/population $P_k$, $1 \leq k \leq n$, represents a standard SEIR unit (as described in the previous section), characterized by a 4-dimensional variable $(S_k, E_k, I_k, R_k)$. The communication between nodes was set up, for consistency, also in a compartmental fashion, so that a fraction of individuals travels through each existing outgoing edge from a specific node to the corresponding adjacent nodes. For simplicity, each node was assumed to have originally $N_0=1000$ individuals. The total number of individuals  $N(t) = \sum [S_k(t) + E_k(t) + I_k(t) + R_k(t)]$ is constant throughout the outbreak (that is, external effects such as travel in and out of the network, birth rate and death rate due to factors independent of the disease were ignored). The corresponding coupled dynamics can then be extended from the system~\eqref{astacio} to the following system:

\begin{eqnarray}
\label{net_system}
\frac{dS_k}{dt} &=& -\beta S_k(I_k+qE_k)/N_0 + q_c CS - q_c \sum_{j} C_{j,k} S_k   \nonumber \\
\frac{dE_k}{dt} &=&  \beta S_k(I_k+qE_k)/N_0 + q_c CE - q_c \sum_{j} C_{j,k} E_k- \delta E_k \nonumber \\
\frac{dI_k}{dt} &=& \delta E_k - \gamma I_k  \nonumber \\
\frac{dR_k}{dt} &=& \gamma I_k 
\end{eqnarray}

\noindent The $4n$ dimensional system~\eqref{net_system} reflects the inner SEIR dynamics of each particular node $P_k = (S_k,E_k,I_k,R_k)$, for $1 \leq k \leq n$, but also captures, in a compartmental way, the population flow between the nodes. The adjacency matrix $C$ is zero diagonal (the graph has no self loops, since travel within one's own community is not relevant). For simplicity, the fraction of travelers $q_c$ between adjacent nodes was taken to be the same for all connected node pairs (and was fixed to $q_c=2.5\%$ in our simulations of this system).

The total number of healthy and respectively latent  individuals leaving node $P_k$ at time $t$, headed towards each of the adjacent nodes, is proportional with the existing healthy and respectively latent population in node $k$. Subsequently, each node will receive an incoming flow of travelers from the nodes connected to it. We based this on the assumption that latent individuals are able to travel out of their own community: being asymptomatic, they have not yet been diagnosed, although they pose a risk in spreading the disease, by adding themselves to the existing latent population at a different node. For obvious reasons, infected individuals cannot travel in our model. In addition, circulation of recovered individuals was ignored, since it would have no effect on the disease dynamics (under the assumption that they can no longer contract or spread the disease, they would only permute between the $R_k$ compartments of different nodes).

\begin{figure}[h!]
\begin{center}
\includegraphics[width=0.4\linewidth]{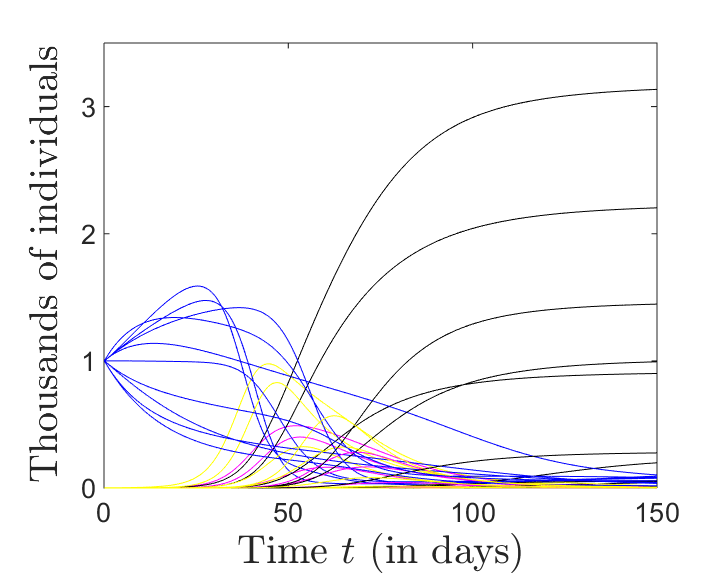}
\includegraphics[width=0.4\linewidth]{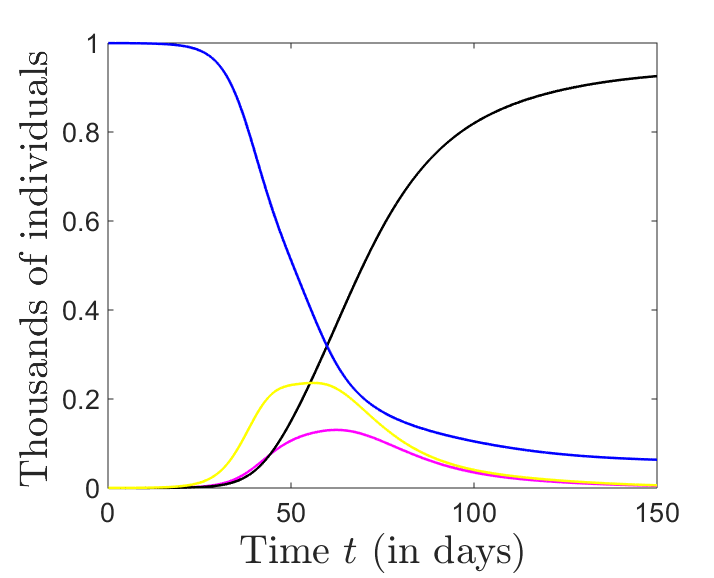}
\end{center}
\caption{\emph{\small {\bf Simulation of interactive dynamics in a network with $n=10$ communities}, in which the contagion was triggered by only one latent individual in $P_1$. {\bf A.} The time evolution of all SEIR variables, obtained by numerically solving system~\eqref{net_system}, is illustrated simultaneously for all $n=10$ nodes, with the $S_k$ shown in pink, the $E_k$ shown in yellow, the $I_k$ shown in red and the $R_k$ shown in black. {\bf B.} The average populations $\overline{S}(t) = \frac{1}{n}\sum_{k=1}^n S(t)$, $\overline{E}(t)$, $\overline{I}(t)$, $\overline{R}(t)$ are shown, with the same color coding.}}
\end{figure}

The model was conceived having in mind typical rural structures, where people travel along established routes between specific locations (for daily or other periodic needs), with the return not necessarily occurring immediately or along the same path (hence the oriented edges). The graph adjacency matrix $C$ delivers a complete description of the communication patterns within the network, and induces instantaneous spread of contagion. Throughout our analysis, we kept all parameters fixed and we focused primarily on how the connection density and patterns affect viral diffusion through the network.

As a start, we studied the effect of edge density on the disease dynamics, in particular on the outbreak impact and duration. For our network model, we define the impact as:

$$R_\infty = \lim_{t \to \infty} \overline{R}(t), \text{ where $\overline{R}(t)$ is the recovered average per community: } \overline{R}(t) = \frac{1}{n} \sum_{k=1}^{n} R_k(t)$$

\begin{figure}[h!]
\begin{center}
\includegraphics[width=0.45\linewidth]{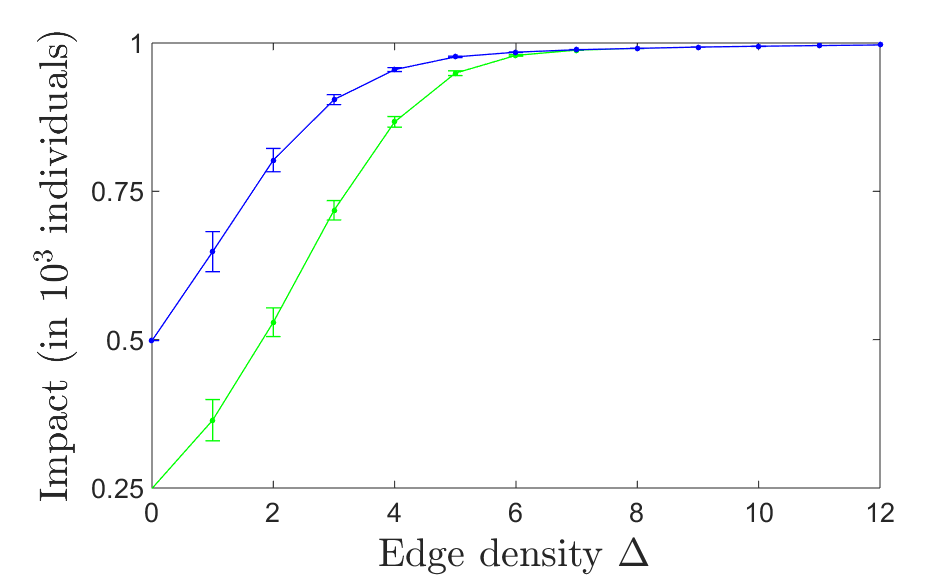}
\includegraphics[width=0.45\linewidth]{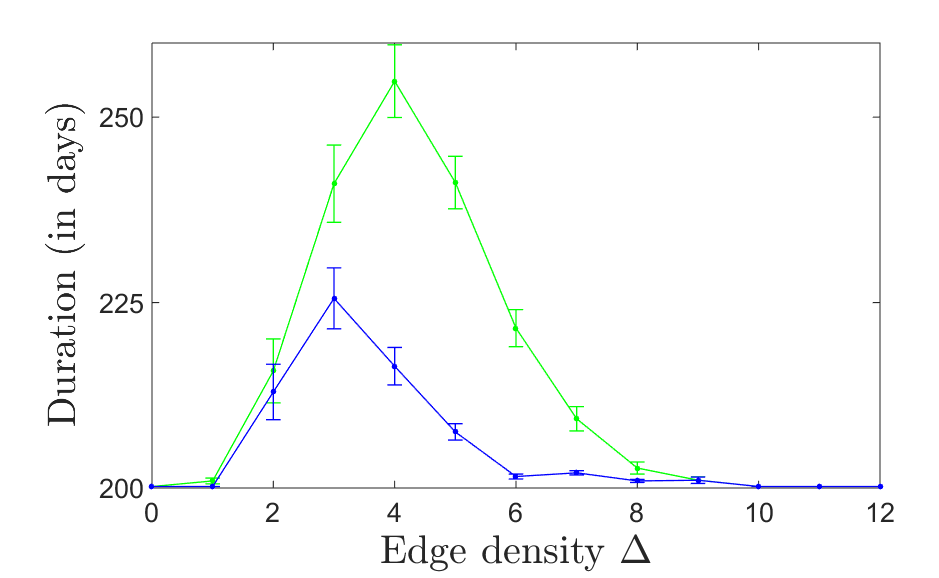}
\end{center}
\caption{\emph{\small {\bf Dependence of the impact $R_\infty$ and of the outbreak duration $L$ on edge density $\Delta$.} For a network of $n=4$ populations each originally with $N_0=1000$ individuals, we illustrate $R_\infty$ and $L$ as the edge density is varied for $\Delta=0$ edges to $\Delta=12$ edges. For each density value, the average impact and duration were computed over all possible $C_{n(n-1)}^\Delta$ adjacency configurations with the respective density. We considered the case of  one focal point consisting of two latent cases, both in the same node (green curve), and the case of two focal points, each consisting of one latent case, starting simultaneously at two distinct nodes (blue curve).} }
\label{connect4_simple}
\end{figure}

\begin{figure}[h!]
\begin{center}
\includegraphics[width=0.45\linewidth]{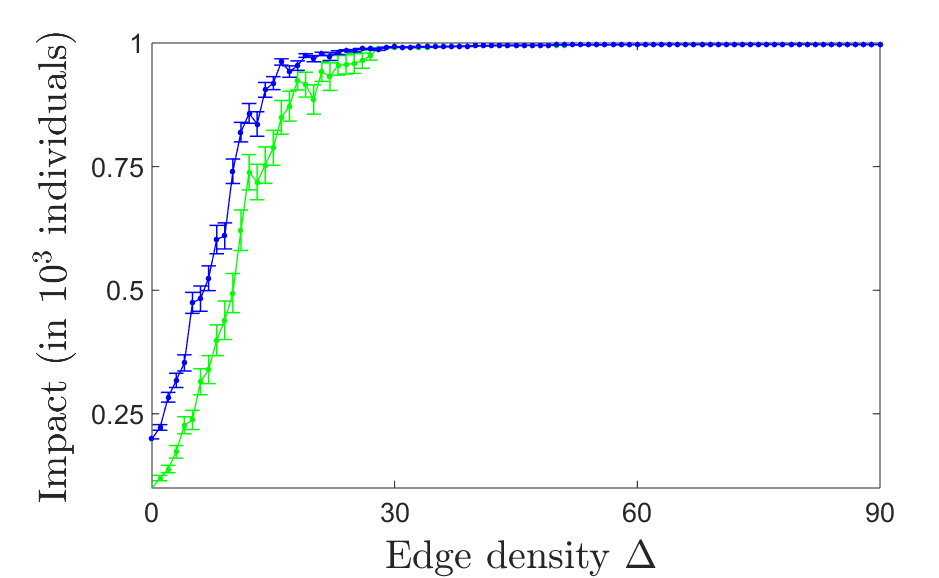}
\includegraphics[width=0.45\linewidth]{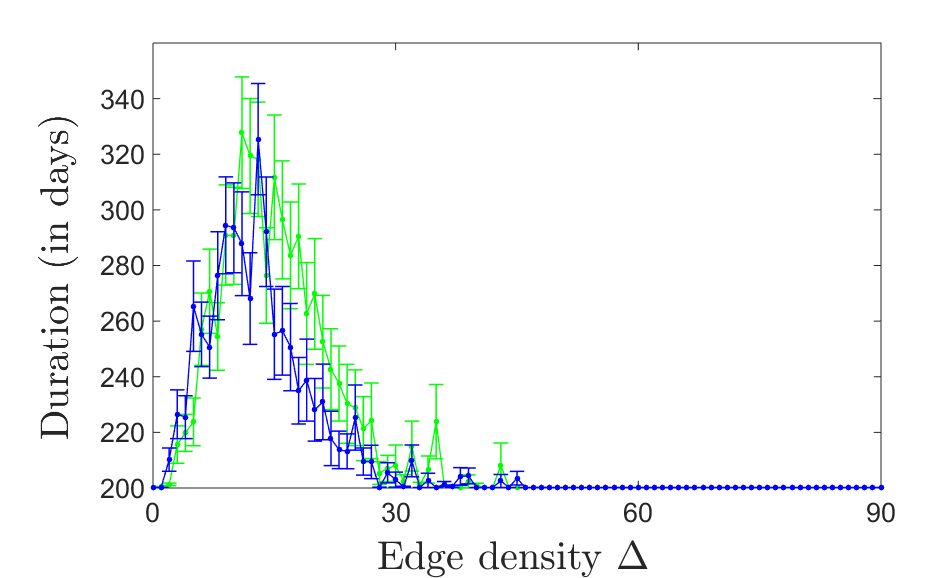}
\end{center}
\caption{\emph{\small {\bf Dependence of the impact $R_\infty$ and of the outbreak duration $L$ on edge density $\Delta$.} For a network of $n=10$ populations each originally with $N_0=1000$ individuals, we illustrate $R_\infty$ and $L$ as the edge density is varied. For each density value, the average impact and duration were computed over a sample of $50$ adjacency matrices chosen at random, with the respective density. We considered the case of  one focal point consisting of two latent cases, both in the same node (green curve), and the case of two focal points, each consisting of one latent case, starting simultaneously at two distinct nodes (blue curve).} }
\label{connect10_simple}
\end{figure}

For this first model, we assume a constant connectivity matrix throughout the process, and we investigate the cases of a single focal point (triggered by two infections), and of two focal points (with one infection each). Fixing the network size, we ran numerical simulations for different connectivity patterns, with the primary aim of investigating the effect of the connectivity density on the outbreak coupled dynamics. For small network sizes ($n \leq 4$), we considered all possible matrix configurations for each fixed edge density $0 \leq \Delta \leq n(n-1)$, and we computed the mean impact over all such configurations (see Figures~\ref{connect4_simple}). For larger $n$'s, we computed a sample-based mean, over a subset of $50$ configurations for each edge density $\Delta$ (Figure~\ref{connect10_simple}). This sample approach was preferred for larger networks, since the number of configurations increases extremely fast with $n$, making a numerical inspection of all configurations for each fixed density very expensive and impractical. In both cases, we plotted the outbreak impact and duration as functions of the edge density  $\Delta$, showing both mean value and error bars (over the distribution of adjacency configurations).

Our computations suggest that, for both one and two foci, the outbreak impact increases with the density of connections, while the duration has a unimodal shape (with a peak in the intermediate connectivity range). This is not a surprising result, and can be explained by the fact that for very low connectivity the contagion is quickly contained, with minimal casualties, and for very high connectivity -- it produces a fast diffusion, with a high number of casualties. It is for intermediate connectivity patterns that the dynamics takes longer to develop. Clearly, one desires to minimize duration, as well as impact (since longer infection times can be as well detrimental for the community). The two functions don't assume their minima in the same conditions, so one may consider optimizing a combination of the two in order to asses the most convenient outcome. It is also worth noticing that the impact has a slower pick-up in the case of one original focus, while the duration is comparable for one or multiple foci in the low connectivity range, then significantly longer for one focus in the intermediate connectivity range. In other words: for intermediate internode communication, the impact is milder if the original infection starts in one single focus, but this scenario loses its advantage in the context of duration.

More precisely: in this preliminary model, $R_\infty$  is relatively low in average only for values of the edge density up to $\sim 20\%$ (as illustrated for $n=4$ in Figure~\ref{connect4_simple} and for $n=10$ in Figure~\ref{connect10_simple}). After crossing a window of very high sensitivity (in our figures, density $15-20\%$), $R_\infty$ increases to its asymptotic value $N_0=1000$ (the total number of individuals per node). This predicts a devastating effect, describing an outbreak that would affect virtually everyone within the linked communities, even at relatively low communication levels. 

Of course, however feared and somber the perspective of a global pandemic, this is not the prediction we expect in reality, based on our existing experience with Ebola outbreaks; hence the model should be revised to reflect more realistic trends. One major flaw of this preliminary model is the assumption of instantaneous and continuous disease spread whenever communication means exist between two nodes. This is clearly not a realistic condition, greatly overestimating the contagion spread, which further reflects in the unrealistically large values of $R_\infty$. Below, we update the model to refine this condition.

Existence of an oriented edge from $k$ to $j$ should only signify that direct communication is \emph{possible} between the two nodes (e.g., there is a road connecting two villages), but not necessarily that this connection is used continuously, or maintained at the same level at all times. We introduce a simple way of accounting for this variability, while keeping the model compartmental (i.e., counting overall transfer rates, rather than keeping track of dynamics of individuals). At each time step $t$, each existing connection from some $k$ to some $j$ is used with a fixed probability $r_{kj}$. For each node, the probability of inward or outward travel can be tuned according to the local or global situation of the network. For example, the flow can be temporarily diminished or cut completely  if the node needs to be quarantined to prevent  infection spread. For our first analysis, we take $r$ for simplicity to be constant throughout the outbreak process, and identical for all outgoing edges. In the following section, we will allow $r$ to adapt, producing a variable distribution of values over all the network edges, changing according to the implementation measures typically taken to minimize the impact of the outbreak.

\begin{figure}[h!]
\begin{center}
\includegraphics[width=0.45\linewidth]{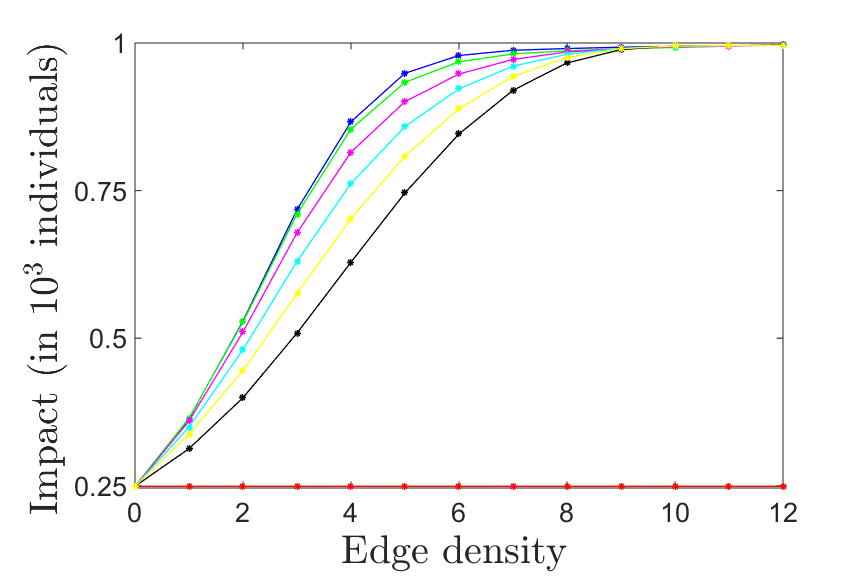}
\includegraphics[width=0.45\linewidth]{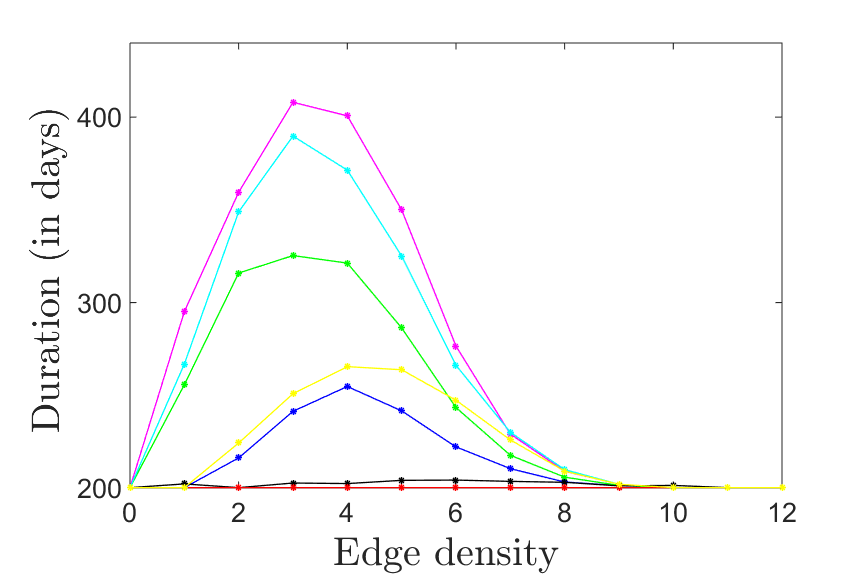}
\end{center}
\caption{\emph{\small {\bf Dependence of outbreak aftermath on edge density $\Delta$ and travel likelihood $r$}, for a network of $n=4$ communities. The impact $R_\infty$ ({\bf A}) and outbreak duration ({\bf B}) are computed, for each fixed edge density (number of 1s in the adjacency matrix) as averages over all the possible configurations of the adjacency matrix, with that fixed density. The different color plots correspond to different values of the probability to travel along an available edge: $r=1$ (blue), $r=0.5$ (green), $r=0.2$ (magenta),  $r=0.1$ (cyan), $r=0.05$ (yellow), $r=0.01$ (black), $r=0$ (red).}}
\label{connect_threshold_4}
\end{figure}

\begin{figure}[h!]
\begin{center}
\includegraphics[width=0.45\linewidth]{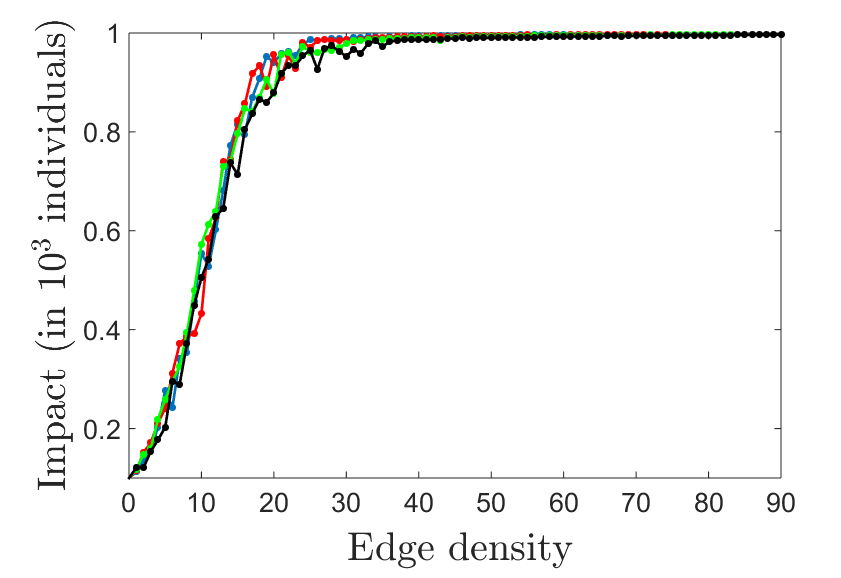}
\includegraphics[width=0.45\linewidth]{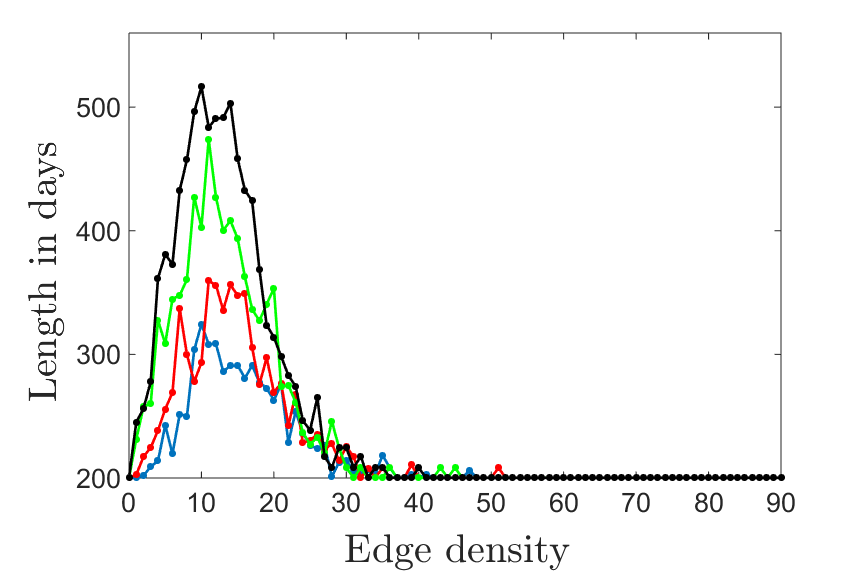}
\end{center}
\caption{\emph{\small {\bf  Dependence of outbreak aftermath on edge density $\Delta$ and travel likelihood $r$}, for a network of $n=10$ communities. The impact $R_\infty$ ({\bf A}) and of the outbreak duration ({\bf B}) are computed, for each fixed edge density, as averages over a sample of $S=50$ adjacency configurations with that fixed density. The different color plots correspond to different values of the probability to travel along an available edge: $r=1$ (blue), $r=0.8$ (red), $r=0.5$ (green). $r=0.2$ (black).}}
\label{connect_threshold_10}
\end{figure}

Figure~\ref{connect_threshold_4}  illustrates in the case of probabilistic travel (described above), and for a single focal point of one individual, the dependence of impact and duration on both the density of edges, and on the probability $r$ to travel out of each node along the prescribed edges. Notice that impact and duration, while having the same qualitative dependence on density, exhibit subtle differences when changing the overall likelihood to travel $r$. For example, in the case of $n=4$ coupled nodes, the impact is $R_\infty = \frac{1}{4} N_0 = 250$ for $r=0$, since no travel implies that only the focal community will be affected (there is no dependence on edge density, since the edges are never used to transport the virus). Moving $r$ to any small positive value acts as a bifurcation, since the effects of the density $\Delta$ appear immediately, causing the impact to increase, more quickly and steeply for larger values of $r$.  These being said, however, the dependence on travel likelihood in the $r>0$ range is not as dramatic in this model as one may expect; it suggests that indiscriminately diminishing travel in the network can't single-handedly accomplish a major decrease in impact, unless the travel is altogether prohibited. 

An even more surprising effect appears when studying the dependence on $r$ of the unimodal curve representing the outbreak duration. While the critical \emph{point} does not vary much -- remaining, for each curve, broadly in the same intermediate density range  for each $n$ (a more precise localization could be obtained at the expense of lengthier computations), the behavior of the critical \emph{value} differs both qualitatively and quantitatively between different $r$ values, first increasing with $r$, and then decreasing. This suggests that indiscriminately lowering $r$ may in fact be detrimental, by increasing the duration of the outbreak, without substantially lowering the impact (especially in the region of intermediate densities). Only a dramatic and implausible shift of $r$ to a value close to zero throughout the network would result in improving both impact and duration. These effects can be observed in both Figures~\ref{connect_threshold_4} and~\ref{connect_threshold_10}, for $n=4$ and $n=10$, respectively.

Altogether it seems clear that, in order to control the outbreak in our model, one cannot just treat the system as a whole, but instead has to target more specific network sites, based on (1) the network architecture and (2) the outbreak's current state throughout the network, starting from the source of contagion.

\subsection{Modular adaptable network and effects of quarantine}

\begin{figure}[h!]
\begin{center}
\includegraphics[width=0.8\linewidth]{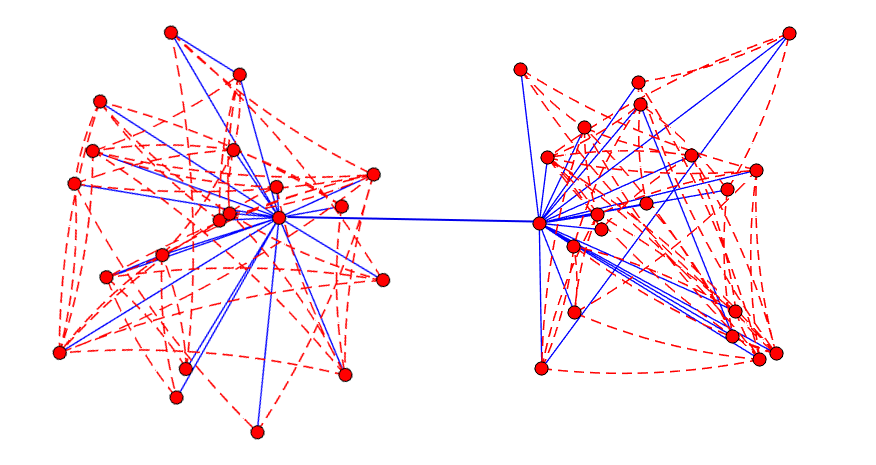}
\end{center}
\caption{\emph{\small {\bf Model network architecture with two interconnected modules}.  Both modules (subgraphs), each with $n=20$ nodes and $\Delta=100$ oriented edges, are organized as hubs, each with one central node connected bidirectionally with all other nodes in the corresponding hub. The two hubs are connected by one single, bidirectional edge between their two central nodes. In this figure, bidirectional edges are represented in blue, and unidirectional edges are represented by dotted red curves, curved counter clock-wise. This type of architecture was used for the simulations, e.g., in Figure~\ref{lag_impact}.} }
\label{hub_network}
\end{figure}

In this section, we will simulate an outbreak in the more realistic context of two interacting subnetworks (or modules), and introduce more structured quarantine measures, with the aim of reducing both impact and duration. As a plausible, but simplified scenario, we consider the modules to be organized as \emph{hubs}, in the sense that each has a central node connected bidirectionally with all other nodes in the respective subnetwork, also allowing a specific number of additional random oriented edges between the other nodes of the respective module (as in the example pictured in Figure~\ref{hub_network}). This aims to represent the interaction of two large structures (e.g., countries), each organized as a network of smaller communities (e.g., towns, or villages). Such a multi-modular graph contains local, intra-modular connections between nodes (e.g., roads between villages), and long-distance, inter-modular connections. In our case we considered, for simplicity, a single inter-modular, bidirectional edge, running between the two central hub nodes (which could be seen as the only significant communication means between the two countries, e.g. -- airports).

We discuss infection transmission and impact in this type of network, and the efficiency of introducing timely quarantines to control the outbreak. We distinguish between two types of quarantine, local (intra-modular) and global (inter-modular), as follows. If  Ebola infection in detected for a period longer than $\theta$ days  in a population/node (i.e., $I_k(t) > 0$ for the specific node $P_k$), a local quarantine is introduced by cutting all in and out connections with the node $P_k$. If infection persists at any node within either hub for longer than $\tau$ days after the local quarantine (with $\tau > 0$), the the two hubs are immediately disconnected (the connecting edge is cut off).  We study how the timing of these isolation measures affects long term dynamics, towards finding a scheme that would minimize the inconvenience of lengthy quarantine, while still delivering efficient outbreak containment. While one naturally expects a prompter quarantine response to lead to a better outcome, our study looks in more detail at the extent to which the timing matters at both local and global levels.

\begin{figure}[h!]
\begin{center}
\includegraphics[width=0.45\linewidth]{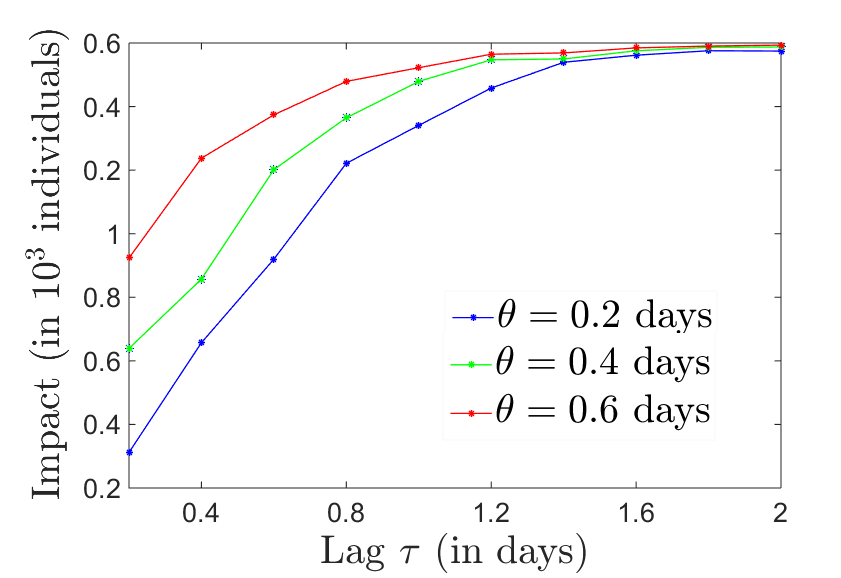}
\includegraphics[width=0.45\linewidth]{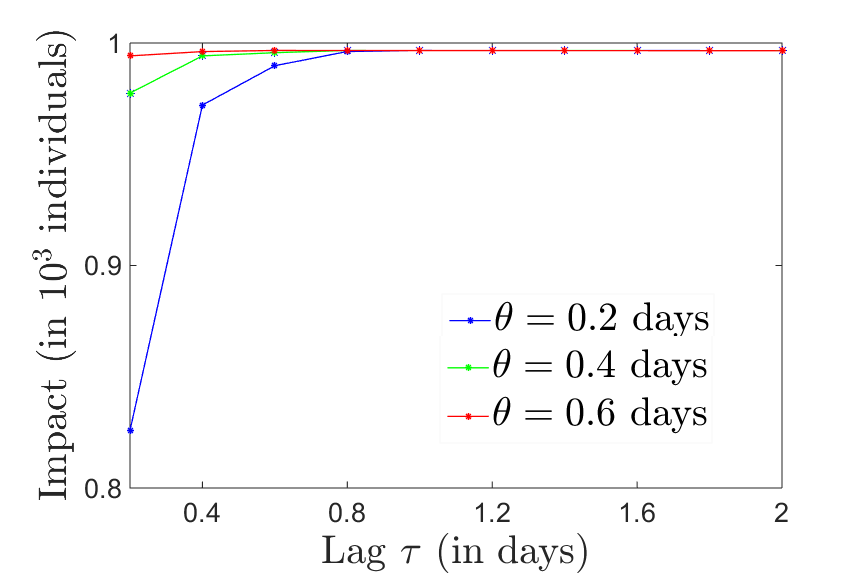}
\includegraphics[width=0.45\linewidth]{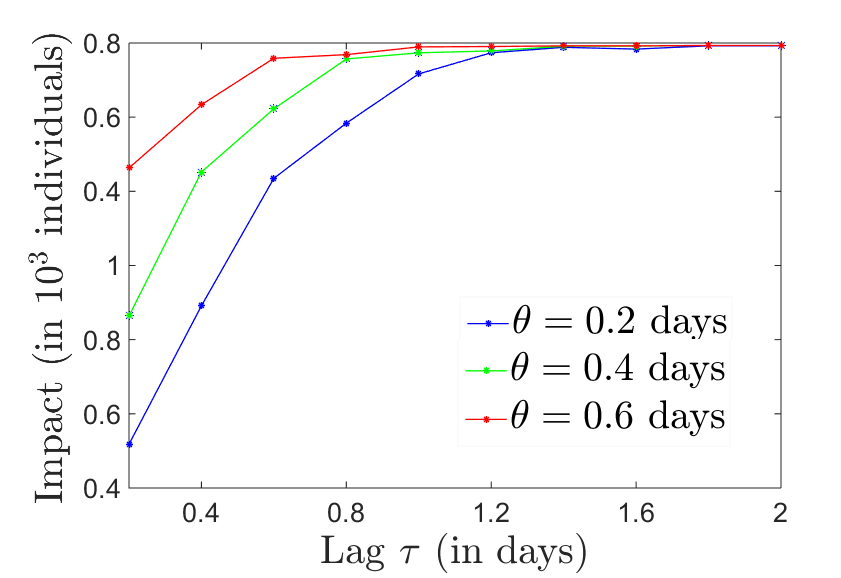}
\includegraphics[width=0.45\linewidth]{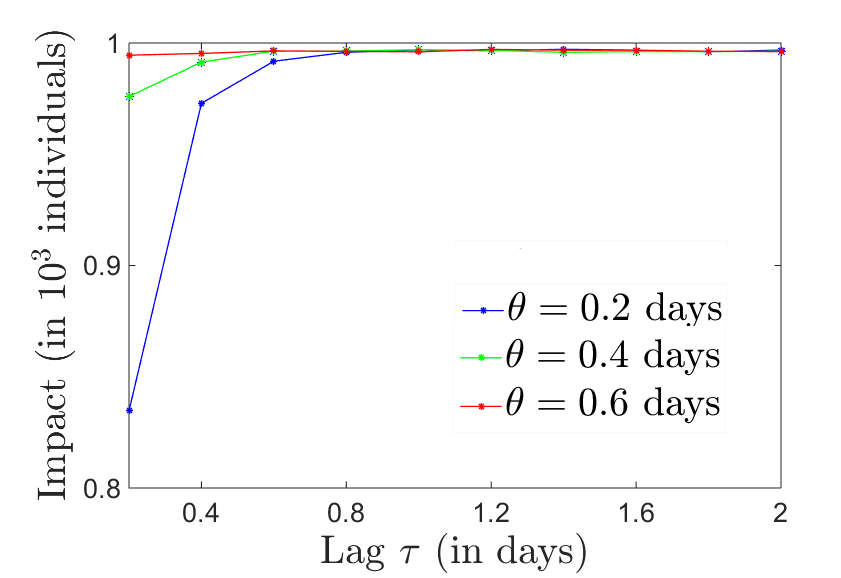}
\end{center}
\caption{\emph{\small {\bf  Dependence of impact $R_\infty$ on quarantine time lags $\theta$ and $\tau$}. Top: for a network composed of two hubs, each with $n=10$ nodes, connectivity density $\Delta = 30$ and {\bf A.} travel likelihood $r_\text{int} = 0.4$ along intra-modular edges and $r_\text{ext} = 0.2$ along the inter-modular edges;  {\bf B.} travel likelihoods $r_\text{int} = r_\text{ext} = 0.5$. Bottom: for a network composed of two hubs, each with $n=10$ nodes, connectivity density $\Delta = 50$ and {\bf C.} travel likelihood $r_\text{int} = 0.4$ along intra-modular edges and $r_\text{ext} = 0.2$ along the inter-modular edges; {\bf D.} travel likelihoods $r_\text{int} = r_\text{ext} = 0.5$. Each curve represents the dependence of $R_\infty$ on the inter-modular lag $\tau$, for a fixed intra-modular lag $\theta$, as follows: $\theta=0.2$ days in blue, $\theta=0.4$ days in green and $\theta= 0.6$ days in red.} }
\label{lag_impact}
\end{figure}

We considered different architectures, edge densities and travel likelihoods ($r_\text{int}$ along the active intra-modular edges and $r_\text{ext}$ along the inter-modular edge), at the start of the outbreak (time of the original infection in the network), and we measured in each case the dependence of the outbreak impact $R_\infty$ on the time delays $\theta$ and $\tau$ of the two quarantines. In Figure~\ref{lag_impact} we illustrate, for modules of size $n=10$, the dependence of $R_\infty$ on $\theta$ and $\tau$ (with each curve showing the dependence on $\tau$ for a different value of $\theta$, increasing from blue to green to red). The critical range (of maximum sensitivity) for $\theta$ is less than one day; the critical range for $\tau$ is a little longer. The top panels reproduce the outcome for a lower connectivity network (30 out of the maximum of 90 possible edges in each module); the bottom panels consider a higher connectivity network (50 edges). The left panels present results for lower travel likelihoods before the infection (intra-modular $r_\text{int}=0.4$ and $r_\text{ext}=0.2$ along the inter-modular edge); the right panels show higher pre-infection travel likelihoods ($r_\text{int}=r_\text{ext}=0.5$).

Generally, as one may expect, for any fixed local delay $\theta$, the impact increases with the inter-modular delay $\tau$ (the longer one waits to completely separate the two modules, the larger the infection-produced damage). Similarly, for each fixed $\tau$, the impact increases with the local delay $\theta$ (with the same wait time before separating the modules, a longer local delay increases the global average damage). However, these qualitative effects present wide quantitative variations with the structure and connectivity of the network.

 One may note that, within the realistic connectivity range, there are no wide differences determined by the edge density \emph{per se} (panels A and C have only subtle quantitative differences, and panels B and D look almost identical); the more substantial difference are induced by people's likelihood to travel along the existing edges before the quarantines are imposed. 

For a network with less traveled edges (left panels), setting early quarantines has a very strong effect (the curves increase steeply for low values of $\tau$, and taper off asymptotically. We noticed, however, that the impact differs a lot with $\theta$, especially for values of $\tau$ less than one day (when the impact gets close to the common asymptotic value). In this case, it seems important to strive for a short local wait $\theta$. However, comparing the rates of change along each curve and across curves, it appears that shortening $\theta$ and $\tau$ has in this case comparable effects on the impact.

For more highly traveled networks (right panels), a slightly large $\tau$ may lead to a catastrophic impact even for very low $\theta$ values. On the other hand, a large $\theta$ single-handedly leads to a dramatic impact, even with very short $\tau$.  In such networks, only a combination of both small $\theta$ and small $\tau$ can substantially reduce the impact. An efficient control of the impact requires  very quick local intervention, followed tightly, possibly even before observing success or failure of local measures, by global separation.

\section{Discussion}

In this paper, we investigated a few aspects of a widely studied theoretical problem: that of dependence of coupled nonlinear dynamics on the architecture of the coupling. We discussed the importance of understanding the mechanics of this problem in the context of disease transmission and control in a population network. For our illustration, we worked with system parameters measured in a historic Ebola outbreak, but the same compartmental construction, the same concepts and methods can be used to study other viral transmissions, or any information diffusion process in a network of coupled nodes.

Aside from common sense conclusions on the necessity of prompt quarantines at early signs of a viral outbreak, our study brings interesting quantitative highlights, with potential applications when customizing and tuning quarantine placement and timing. There are a few particular observations that may be of general value when assessing the importance of quarantine and other measures to control an outbreak. First, we saw that conditions which minimize the outbreak impact may not be ideal for other aspects (e.g., duration), as important for the affected population. Second, a prompt local quarantine is generally helpful, but cannot efficiently lower the impact in and off itself, if not paired with a prompt  global separation. A delay in global separation may make the efficiency of local quarantine irrelevant. Third, fast global separation is optimal if paired with a prompt local quarantine, and in some cases this can efficiently and substantially lower the impact.

This suggests that a helpful approach to infection control in a hypothetical outbreak would require authorities to asses the type of network and travel flow that is at risk, and tune the quarantine timing to occur within the respective optimal ranges. For example, in the case of a highly traveled modular network, the global separation may have to be imposed faster than common sense suggests, even if this may involve additional resources in putting in motion the slow global machinery.

Our study is only one step, among a few others~\cite{kopman2012epidemicsim,valdano2015analytical,siettos2015modeling}, towards understanding the importance of the architecture and hardwiring in a network exposed to an epidemic outbreak. It has clear limitations, but also a wide potential for extensions (to other disease dynamics, other population networks or even more general dynamic networks), and for more elaborate analyses of the underlying mathematical phenomena. Some of these aspects are discussed below.

One limitation introduced conceptually in the model relates to the strict assumptions made on Ebola spread and immunity. For example, we worked from the start under the hypothesis that the individuals who had contracted Ebola once, cannot have the disease again. This was based on a corresponding assumption in our original reference paper~\cite{astacio1996mathematical}, which in turn was supported by the lack of evidence of any individuals with more than one Ebola infection within their life span. This idea is currently considered controversial in the disease dynamics literature, especially with the known variety of Ebola virus strains which may make a prior infection with one strain irrelevant immunologically to a new infection with a different strain. Immunity built-up aspects, such as duration and effectiveness to multiple viral strains, have been also explored mathematically~\cite{piazza2013bifurcation}.

Another simplifying assumption we made was that infected people can no longer spread the virus after the infection clears, whether this occurs through recovery or death of the individual. In reality, this may not be accurate. While some studies show that the risk of transmission from bodily fluids of convalescent patients is low (when infection control guidelines for the viral hemorrhagic fevers are followed ~\cite{bausch2007assessment}), other studies have shown that the Ebola virus can be spread through the sperm of recovered individuals for up to 7 weeks after infection~\cite{ahmad2014ebola}. It is also well known at this point that dead bodies can remain contagious for up to 60 days~\cite{ahmad2014ebola}, with the potential of infection spread though contact with the dead body (e.g., during ritual funerals~\cite{chippaux2014outbreaks}). Further iterations of the model may consider introducing these effects into the coupled equations.

Finally, one important aspect to explore in future studies is the extent to which details of the network configuration can flip the optimal quarantine circumstances. The current paper shows that the quarantine measures required for maximal control may differ with measures such as edge density, or travel likelihood. Other studies have explored the impact of the small-worldness, or assortativeness of the network on the overall dynamics~\cite{kopman2012epidemicsim}. There is, however, the likely possibility that there is no canonically optimal response based only on global network measures, and that an adequate quarantine systems has to be customized in response to the local details of the network architecture. Returning to out original analogy between viral diffusion and brain dynamics -- in the same fashion in which clinical neuroscience is evolving towards ``brain profiling,''  and personalized clinical assessments, in the same way the response to a global Ebola outbreak may have to consider a ``population network profiling'' in order to deliver optimal results. 

Our small values for optimal quarantine time lags suggest that full preparedness for a global outbreak may involve having a pre-established plan of action, to avoid fatal computation and decision-related delays during the spreading of the outbreak. This would require constantly updated knowledge of local and global travel patterns, a dynamic ``global connectivity map'' that could be implemented directly, when necessary, into simulations, and produce immediate predictions and provide efficient choices for quarantines. The cost of maintaining online global information may be a well placed investment in view of a potential pandemic.

\bibliographystyle{plain}
\bibliography{references}

\end{document}